\documentclass[aps,prl,twocolumn,showpacs,showkeys,superscriptaddress]{revtex4}

\usepackage{amsmath}
\usepackage{amssymb}
\usepackage{graphicx}
\usepackage{dcolumn}
\usepackage{bm}
\begin{document}

\title{Optical properties of graphene nanoribbons: 
the role of many-body effects}

\author{Deborah \surname{Prezzi}}
\email[corresponding author: ]{prezzi.deborah@unimore.it}
\affiliation{INFM-CNR-S3, National Center on nanoStructures
             and bioSystems at Surfaces, I-41100 Modena, Italy.}
\affiliation{Dipartimento di Fisica, Universit\`a di
             Modena e Reggio Emilia, I-41100 Modena, Italy.}
\author{Daniele \surname{Varsano}}
\affiliation{INFM-CNR-S3, National Center on nanoStructures
             and bioSystems at Surfaces, I-41100 Modena, Italy.}
\author{Alice \surname{Ruini}}
\affiliation{INFM-CNR-S3, National Center on nanoStructures
             and bioSystems at Surfaces, I-41100 Modena, Italy.}
\affiliation{Dipartimento di Fisica, Universit\`a di
             Modena e Reggio Emilia, I-41100 Modena, Italy.}
\author{Andrea \surname{Marini}}
\affiliation{Dipartimento di Fisica, Universit\`a di
             Roma "Tor Vergata", I-0133 Roma, Italy.}
\author{Elisa \surname{Molinari}}
\affiliation{INFM-CNR-S3, National Center on nanoStructures
             and bioSystems at Surfaces, I-41100 Modena, Italy.}
\affiliation{Dipartimento di Fisica, Universit\`a di
             Modena e Reggio Emilia, I-41100 Modena, Italy.}
\date{\today}

\begin{abstract}
\noindent
We investigate from first principles the optoelectronic properties
of nanometer-sized armchair graphene nanoribbons (GNRs). We show that
many-body effects are essential
to correctly describe both energy gaps and optical response.
As a signature of the confined geometry, we observe
strongly bound excitons dominating the optical spectra,
with a clear family dependent binding energy.
Our results demonstrate that GNRs constitute 1D
nanostructures whose absorption and
luminescence performance can be controlled by changing
both family and edge termination.
\end{abstract}

\maketitle

Graphite-related nanoscale materials, such as fullerenes and nanotubes, have
long been the subject of an intense research for their remarkable 
properties~\cite{dres+96book}. 
The recent discovery of stable, single-layer 
graphene~\cite{novo+05nat,zhan+05nat,berger} has prompted the attention on a 
different graphitic quasi-1D nanostructure, i.e. graphene nanoribbons (GNRs).
These systems have been theoretically studied in the past 
decade~\cite{fuji+96jpsj,naka+96prb,waka+99prb,kawa+00prb,kusa-maru03prb}
as simplified models of defective nanotubes and 
graphite nano-fragments. 
However, only very recently isolated nanometer-sized GNRs 
have been actually synthetized by etching larger 
graphene samples, or by CVD growth on suitably patterned 
surfaces~\cite{chen+07condmat,han+07condmat,tana+02ssc}.  
The production techniques advanced in these pioneering
works are expected to become highly controllable, opening up new avenues 
for both fundamental nanoscience and nanotechnology applications.

One of the most striking features of GNRs is the high sensitivity of 
their properties to the details of the atomic structure~\cite{fuji+96jpsj,
naka+96prb,waka+99prb,ezaw06prb,baro+06nl,son+06prl,pisa+07prb}. 
In particular, the edge shape dictates their classification in armchair 
(A), zigzag (Z) or chiral (C) ones, thus determining 
their energy band gaps. In addition to an overall decrease of energy gaps with 
increasing ribbon width, also observed experimentally~\cite{han+07condmat}, 
theoretical studies predict a superimposed oscillation feature
~\cite{baro+06nl,ezaw06prb,son+06prl}, which is maximized for A-GNRs. 
According to this behaviour, A-GNRs are further classified in three distinct 
families, i.~e. $N=3p-1$, $N=3p$, $N=3p+1$, with $p$ integer, where $N$ 
indicates the number of dimer lines across the ribbon width. 
This fine sensitivity to the atomic configuration raise the opportunity 
to tailor the optoelectronic properties of A-GNRs
by appropriately selecting both ribbon family and width.

In spite of this interest, previous theoretical studies of the electronic 
 (see e.g. Refs.~\onlinecite{son+06prl,pisa+07prb,naka+96prb}) 
and optical properties~\cite{baro+06nl} of GNRs were only based on the 
independent-particle approximation or on semi-empirical calculations.
However, many body effects are expected to play a key role in low 
dimensional 
systems~\cite{ruin+02prl,rohl-loui99prl,chan+04prl,spat+04prl,brun+07prl}
due to enhanced electron-electron correlation.
Motivated by this theoretical issue and by recent
experimental progress~\cite{chen+07condmat,han+07condmat,tana+02ssc}
pursuing the potential of GNRs for nanotechnolgy applications,
we have carried out {\it ab initio} calculations to study the effects of
many-body interactions on the optical spectra of 1-nm-wide A-GNRs 
belonging to different families.

In this Letter, we show that 
a sound and accurate description of the 
optoelectronic properties of A-GNRs must include many-body effects.
We will demonstrate that 
there are many signatures of the non-local correlations occurring in these 
confined systems. First of all, quasiparticle corrections are
found to be strongly state-dependent. Moreover, 
the optical response of A-GNRs is dominated by prominent 
excitonic peaks, 
with a complex bright-dark structure which would not
have been even expected from an independent-particle framework.
Both quasi-particle corrections and exciton binding energies 
are found to exhibit
an oscillating behaviour, according to the family classification.
Finally, the electronic and optical properties of hydrogen passivated A-GNRs
are compared with those of clean-edge ribbons: including 
many-body effects allows us to single out the impact of this edge
modification on absorption and luminescence.

The first-principles calculation of the optical excitations is carried
out using a many-body perturbation theory approach, based on a three-step
procedure~\cite{note-review}.
As a preliminary step, we obtain the ground state electronic properties of 
the relaxed system, by performing a density-functional theory (DFT)
supercell calculation,
within the local density approximation (LDA)~\cite{Pwscf,note-dft}. 
Second, the quasiparticle corrections to the LDA eigenvalues are evaluated
within the $G_0W_0$ approximation for the self-energy operator, 
where the LDA wavefunctions are used as good approximations for the 
quasiparticle ones, and the screening is treated within the plasmon-pole 
approximation~\cite{godb-need89prl}. 
Third, the electron-hole interaction is included by solving the 
Bethe-Salpeter (BS) equation in the basis set of quasielectron and 
quasihole states, where the static screening in 
the direct term is calculated within the random-phase approximation (RPA).
Only the resonant part of the BS hamiltonian is taken into account
throughout the calculations (Tamm-Dancoff approximation), since   
we have verified that the inclusion of the coupling part  
does not affect significantly the absorption spectra~\cite{note-coupl}. 
Moreover, only the case of light polarized along the ribbon axis is examined, 
as a significant quenching of optical absorption is known to occur in 1D 
systems for polarization perpendicular to the principal axis~\cite{mari+03prl}. 
All the $GW$-BS calculations are performed with the code SELF
~\cite{Self,note-bse}.

To treat an isolated system in the supercell approach, we consider a
separation of 40 a.u. between images in the directions perpendicular to
the ribbon axis. Moreover, in both $GW$ and BS calculations, we truncate
the long-range screened Coulomb interaction between periodic images, in
order to avoid non-physical interactions~\cite{rozz+06prb}.
Due to the rectangular geometry of the system, we use a box-shaped
truncation~\cite{vars-mari07unp}.


\begin{figure}[t]
\includegraphics[width=.40\textwidth,clip]{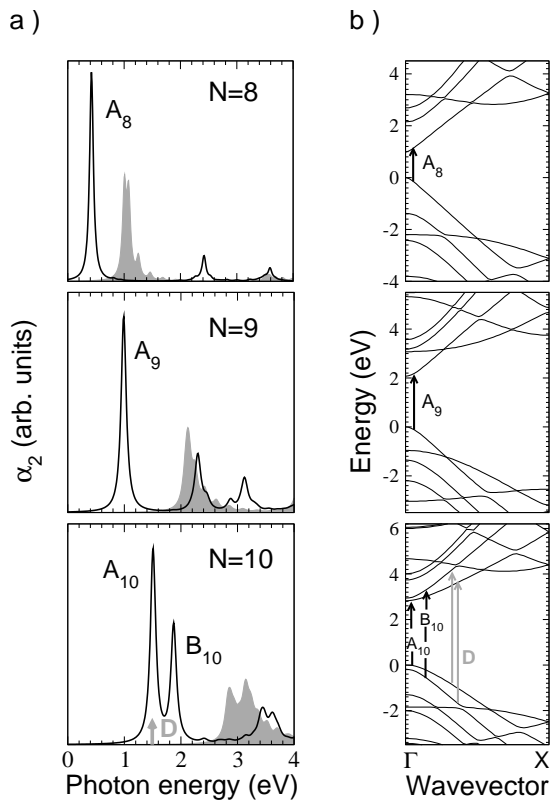}
\caption{ 
(a) Optical absorption spectra of 1 nm wide hydrogen-passivated GNRs:
$N=8$ (1.05 nm wide), $N=9$ (1.17 nm) and $N=10$ (1.29 nm).
In each panel, the solid line represents the spectrum with electron-hole 
interaction, while the spectrum in the single-particle picture is in grey. 
All the spectra are computed introducing a lorentzian broadening.
(b) Quasiparticle bandstructures.
}\label{spectra-band}
\end{figure}

\begin{figure}[bh]
\includegraphics[width=.45\textwidth,clip]{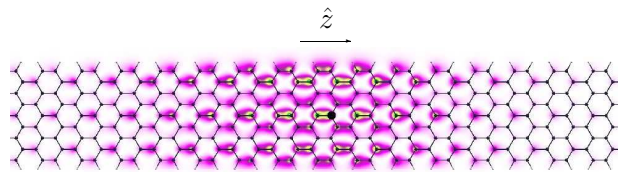}
\caption{
In-plane spatial distribution of the electron for a fixed hole
position (black dot), corresponding to the lowest excitonic peak in the 
$N=9$ case. The spatial density is averaged over the direction orthogonal 
to the ribbon plane.
Dimension of the panel: $1.2 \times 6.4$ nm.
}\label{exc-wfc}
\end{figure}

%

We start by considering 1 nm wide hydrogen-passivated A-GNRs belonging to 
different families, namely $N=8$, 9, 10. 
Figure~\ref{spectra-band}~(a) depicts their calculated optical absorption 
spectra, while the quasiparticle bandstructures are shown in
Fig.~\ref{spectra-band}~(b). All the results are summarized in
Table~\ref{tab}. 
The quasiparticle $GW$ corrections open the LDA energy gaps at $\Gamma$
by 0.72, 1.32 and 1.66 eV for $N=8$, 9 and 10, respectively. 
These energy corrections are larger than those of bulk semiconductor with 
similar LDA gaps, due to the enhanced Coulomb interaction in low dimensional
systems. 
In addition, a family modulation of the corrections can be noticed, with 
larger corrections for the GNRs with larger LDA gaps.  
The gap opening is accompanied by an overall stretching of the banstructure 
of $17-22\%$, similar to the value found for graphene 
(about $20\%$)~\cite{angel}. 

In the absence of e-h interaction, such a bandstructure would result in the 
optical absorption spectra depicted in grey [Fig.~\ref{spectra-band}~(a)], 
characterized by prominent 1-D van Hove singularities. 
The inclusion of the excitonic effects (solid black line) dramatically modifies
both the peak position and absorption line-shape, giving rise to individual 
excitonic states below the onset of the continuum, with binding energy 
of the order of the eV.

The lowest-energy absorption peaks for $N=8$ and 9, labelled $A_8$ and $A_9$, 
have the same character: in both cases, the principal contribution comes from 
optical transitions between the last valence and first conduction bands, 
localized in k-space near the $\Gamma$ point [Fig.~\ref{spectra-band}~(b)].
The binding energies for these lowest optically active excitons are 
0.58 and 1.11 eV for $N=8$  and 9, respectively.
As compared to the first two systems, the $N=10$ GNR shows a richer low-energy 
spectrum. 
Each noninteracting peak gives rise to a bright excitonic state [arrows 
$A_{10}$ and $B_{10}$ in Fig.~\ref{spectra-band}~(b)], with binding 
energies of 1.31 and 0.95 eV. In addition, the mixing of dipole forbidden 
transitions between the same bands [arrows $D$ in Fig~\ref{spectra-band}~(b)] 
is responsible for an optically inactive exciton degenerate in energy 
with $A_{10}$. The $D$ state thus provides a competing path for non radiative 
decay of optical excitations,
which could affect the luminescence yield of the system.   
This feature results from transitions between pairs of bands very close 
in energy to each other, and is therefore expected to be a common outcome 
for all $N=3p+1$ GNRs.  

A further insight in the effects of electron-hole interaction is provided by 
the evaluation of the resulting spatial correlations. 
In Fig.~\ref{exc-wfc}, we plot the in-plane 
probability distribution of the electron for a fixed hole position (black dot),
corresponding to the lowest excitonic state in the $N=9$ case. 
While the electron distribution extends over the whole ribbon width, 
the modulation of the exciton wavefunction $|\psi({\bf r_e}; {\bf r_h})|^2$ 
along the ribbon axis is 
entirely determined by the Coulomb interaction. 
Similar wavefunctions (not reported here) for the lowest excitons have been 
obtained for GNRs of different families, with spatial extentions
~\cite{note-ext} of about 34, 23 and 18 \AA~for $N=8$, 9 and 10, respectively.

\begin{table}[t!]
\begin{ruledtabular}
\begin{tabular}{lcccc}
 $N$   & LDA & $GW$ & BS & $E_b$ \\
\hline
 8-H & 0.28 & 1.00 & 0.42 &  0.58\\
 8   & 0.50 & 1.59 & 0.71 &  0.88\\
 9-H & 0.78 & 2.10 & 0.99 &  1.11\\
 9   & 0.56 & 1.50 & 0.64 &  0.86\\
10-H & 1.16 & 2.82 & 1.51, 1.87 & 1.31, 0.95 \\
10   & 1.09 & 2.64 & 1.46, 1.68 & 1.18, 0.96 \\
\end{tabular}
\end{ruledtabular}
\caption{ Energy gap (2nd and 3rd columns) and peak position (4th column)
for $N=8$, 9 and 10 A-GNRs, with (-H) and without hydrogen passivation of
the edge sites. The relative binding energies are reported in the last
column. All the values are in eV.
}\label{tab}
\end{table}

%
\begin{figure}[t]
\includegraphics[width=.42\textwidth,clip]{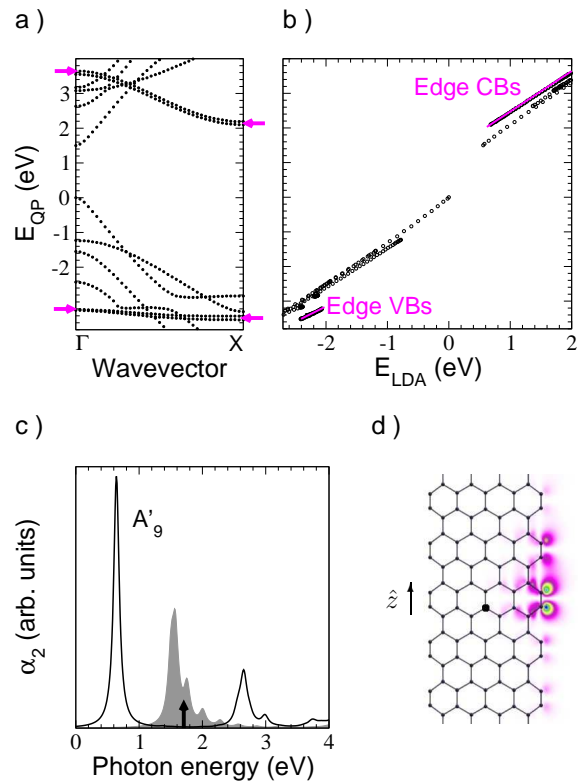}
\caption{
(a) Quasiparticle bandstructure of the $N=9$ hydrogen-free GNR. Arrows
indicate the edge related single particle bands.
(b) Plot of the $GW$ quasiparticle energies vs the LDA energies. 
(c) Optical absorption spectrum, with (solid black) and without 
(grey) excitonic effects. The black arrow indicates the energy 
position of the optically forbidden edge-related exciton. 
Its excitonic wavefunction is depicted in panel (d), whose dimension is 
1.0$\times$2.2 nm. 
}\label{clean}
\end{figure}

We now consider the case of clean-edge nanoribbons, since 
this simple variation of the structure has been often suggested for ribbons
obtained by high-temperature treatments or by dehydrogenation of
hydrocarbons~\cite{kawa+00prb,baro+06nl,rade-bock05jac}. 
This analysis allows us to further explore 
the role played by edge effects in the optoelectronic properties.
Our results are summarized in Fig.~\ref{clean} and Table~\ref{tab}.
As expected, the hydrogen removal leads to a major edge reconstruction,
with the appearence of carbyne-like structures. In fact, the bond 
length for the edge dimers 
reduces from 1.36 for the passivated ribbons to 1.23 \AA~for the clean ones,
pointing to the formation of C-C triple bonds at the edges.
This edge modification leads to a variation of the energy gaps,
such that the distinction between $N=3p-1$ and $N=3p$ families vanishes,
in agreement with previous results~\cite{baro+06nl}.

In Fig.~\ref{clean}~(a), we report the quasiparticle bandstructure for the 
$N=9$ bare ribbon. The main difference with respect to its passivated 
counterpart is the presence of edge-related bands (see arrows) 
in the low-energy optical window. Hence, we focus our attention on the 
properties of these edge states and their influence on the optical response.
These states show the same energy dispersion and real-space 
localization, irrespective of both family and size, already in the
LDA framework~\cite{note-check}: due to this independence on 
bulk properties, their presence is reasonably expected for all 
non-passivated ribbons.  
The self-energy corrections to the LDA eigenvalues are similar to those of 
the passivated systems for the $\pi$ and $\pi^*$ bulk states. The 
edge states show quite a different correction, being deeper in energy 
and with a smoothed stretching with respect to the other bands 
[Fig.\ref{clean}~(b)]. 
This behaviour is to be ascribed to the different degree of real-space
localization between bulk and edge states, and it can be 
singled out by virtue of the non-local character of the self-energy operator
in the $GW$ framework, which is not correctly described
within LDA. 

The aformentioned modification of the bandstructure results in a 
correspondent blueshift ($N=8$) or redshift ($N=9$) of the lowest excitonic 
peak, with $A^{'}_8$ and $A^{'}_9$ becoming almost degenerate, with binding 
energies of about 0.9 eV. 
For the case of $N=10$, we find an inversion of the first and second 
conduction bands, which results in the $B^{'}_{10}$ peak lying below 
$A^{'}_{10}$ and $D^{'}$ almost degenerate in energy with 
$B^{'}_{10}$.
In addition, the edge states introduce an optically inactive exciton, which 
arises from transitions among several bulk valence bands and the conduction 
edge states over the whole Brillouin zone.
This {\it edge exciton} is present in all the studied nanoribbons and is 
located at about 1.4-1.7 eV (black arrow in Fig~\ref{clean}~(d)), with very 
little dependence on family and size ~\cite{note-check}.
This results in the edge exciton being above the first excitonic peak for $N=8$
and 9, and between the first and the second peaks for $N=10$.
We remark that the accurate evaluation of quasi-particle corrections
within $GW$, i.e. beyond the usual approximation based on a uniform band 
stretching on top of a rigid energy shift, is crucial to determine the exact 
energy position of the dark edge excitons relative to the bright ones.

To better understand the character of the edge-related dark state, we plot
its excitonic wavefunction for the case $N=9$ in Fig.~\ref{clean}~(d). 
The mixing of transitions over the whole Brillouin zone induces a strong 
localization of the edge exciton along the ribbon axis, with an extent of only 
$\sim 5$ \AA, that is 4-7 times smaller than the Wannier-like 
{\it bulk excitons} (see Fig.~\ref{exc-wfc}). 

In summary, we have found that the analysis of
the electronic and optical features of GNRs requires 
a state-of-the-art approach within the
many-body perturbation theory, and beyond the DFT framework.
Many-body effects reveal that nanosized A-GNRs
retain a quasi-1D character,
which induces the suppression of the van Hove singularity,
typical of non-interacting 1D systems, and the appearence
of strong excitonic peaks in the optical absorption spectrum.
The lowest excited states
in GNRs are Wannier-like excitons and their binding
energy as well as their luminescence properties
are strongly dependent on the ribbon family.
We investigate the role of many-body effects on the edge-states
arising in 
non-passivated GNRs: our analysis could provide a practical tool for
revealing the nature of the edges in realistic samples.
We demonstrate that GNRs are intriguing systems with tunable 
optoelectronic features, that we quantitatively evaluate
through our calculations. 
The present study  calls for experiments addressing the
optical response of GNRs:
A combined theoretical and experimental understanding
of ribbon size, family and edge-termination
as control parameters for their performance 
can be considered as the first step towards the design
of graphene-based applications
in nanoscale optoelectronics.

We are grateful to A. Rubio, A. C. Ferrari, S. Piscanec, B. Montanari, 
T. Weller, M. Rontani and C. Cavazzoni for stimulating
discussions. We acknoweledge CINECA CPU time granted through INFM-CNR.
D. V. and A. M.  thank the European Nanoquanta NoE 
(NMP4-CT-2004-500198) and the European Theoretical Spectroscopy Facility 
(ETSF).



\begin{thebibliography}{10}
\bibitem{note-dft}
The DFT-LDA calculations are carried out using separable norm-conserving
  pseudopotentials and a plane-wave basis set. A kinetic energy cutoff of 50 Ry
  is employed. Each atomic structure has been fully relaxed, until forces
  acting on atoms were less than 0.01 eV/\AA. The relaxed configurations are
  always planar, for both passivated- and clean-edge structures.
\bibitem{note-coupl}
We have calculated the excitation energies with and without the inclusion of 
  the coupling part for few cases and always found differences well below our 
  computational accuracy (0.1 eV).
\bibitem{note-bse}
All $GW$-BS calculations have been converged with respect to the number of 
  conduction bands, the k-points sampling, the cutoff on local field effects, 
  until errors were less than 0.1 eV.
\bibitem{note-ext}
We calculate the extention as the full width at half maximum of the probability
  distribution.
\bibitem{note-check}
We check the results for few other cases up to 2 nm.
\end{thebibliography}
\end{document}